\documentclass[aps,prl,twocolumn,nobibnotes,superscriptaddress,floatfix,tightenlines,showpacs,notitlepage]{revtex4-1}
\usepackage{graphicx}
\usepackage{bm}
\usepackage[normalem]{ulem}
\usepackage{amsfonts}
\usepackage{color}
\usepackage{ulem}
\usepackage{amsmath}    
\usepackage{epsfig}
\usepackage{subfigure}  
\usepackage[breaklinks,colorlinks = true,linkcolor = blue,urlcolor=blue,citecolor=blue]{hyperref}
\usepackage{amssymb}
\usepackage{lineno}
\usepackage{hyperref}

\newcommand{\pik}{\Pi(k)}

\usepackage[dvipsnames]{xcolor}

\newcommand{\oist}{Complex Fluids and Flows Unit, Okinawa Institute of Science and Technology Graduate University, Okinawa 904-0495, Japan}
\newcommand{\icts}{International Centre for
  Theoretical Sciences, Tata Institute of Fundamental Research,
  Bangalore 560089, India}
\newcommand{\mpi}{18/291A Puthenpurackal, Kottayam 686564, Kerala, India}
\begin{document}
\title{Intermittency, fluctuations and maximal chaos in an emergent universal state of active turbulence}
\author{Siddhartha Mukherjee}
\email{siddhartha.m@icts.res.in}
\affiliation{\icts}%
\author{Rahul K. Singh}
\email{rksphys@gmail.com}
\affiliation{\icts}%
\affiliation{\oist}
\author{Martin James}%
\email{martin.james@yahoo.com}
\affiliation{\mpi}%
\author{Samriddhi Sankar Ray}%
\email{samriddhisankarray@gmail.com}
\affiliation{\icts}%

\begin{abstract}

A hydrodynamic model of active, low Reynolds number suspensions, shows the
	emergence of an asymptotic state with a universal spectral scaling and
	non-Gaussian (intermittent) fluctuations in the velocity field. Such
	states arise when these systems are pushed beyond a critical level of
	activity and show features akin to high Reynolds number, inertial
	turbulence. We provide compelling numerical and analytical evidence for
	the existence of such a \textit{transition} at a critical value of
	activity and further show that the maximally chaotic states are tied to
	this transition.

\end{abstract}

\maketitle


The emergent fluid behaviour of a two-dimensional, dense suspension of motile
bacteria~\cite{wensink2012meso,marchetti2013hydrodynamics,ramaswamy2017active,
alert2021} is susceptible to a wide range of dynamical
phases~\cite{james2021emergence,Sid21,puggioni2022giant}. This makes such systems and their 
characterisation quite distinct from our more accustomed understanding of
(classical) inertial fluids which typically undergo a laminar-turbulence
transition at moderately large Reynolds numbers~\cite{Frisch-CUP}. Amongst the
different dynamical phases, the \textit{active turbulence}  regime is arguably
the most vexing~\cite{alert2021,bhattacharjee2022activity}. While the suspensions are decidedly low
Reynolds number, this phase displays features which seem to suggest that an
analogy with inertial, high Reynolds number turbulent flows is not entirely out of
place~\cite{wensink2012meso}. Yet, the question of whether low Reynolds number
\textit{active flows} can truly be considered turbulent and if indeed the
physics is universal for such systems remains to be fully answered. These are of course important questions
not only from the point of view of theoretical, non-equilibrium physics but
also from a biological perspective. The underlying reasons for why nature
allows such complex, emergent flows in a suspension of active agents, with
reasonably simple rules of interactions and motion, ought to be intrinsically
related to optimal strategies for evasion and foraging~\citep{humphries2010environmental,volpe2017topography}. Recent studies have
shown that Lagrangian measurements underline a key distinction between these
two classes of turbulence whose origin perhaps lies in what is best for the
microorganisms which constitute such flows~\citep{Sid21,Sid22}. However, in the
active turbulence phase, does there exist a limiting behaviour upon increasing activity and a universal state similarly to inertial turbulence with the
Reynolds number going to infinity? In particular, are the tell-tale signatures
of inertial turbulence, namely (approximate) scale-invariance
with a universal spectral scaling exponent, fluctuations, intermittency, and chaos
replicated in active turbulence?
 
In this paper we address these questions and show that for values of
activity beyond a critical threshold, which are nevertheless consistent with velocities in experimentally
realisable bacterial flows~\cite{wensink2012meso,sokolov2012physical},
fluctuations of the velocity field are intermittent (non-Gaussian) 
and accompanied by a scale-invariant distribution of kinetic energy across
Fourier modes with a universal scaling exponent. The existence of such a critical activity---at which a universal, 
turbulent and maximally chaotic state emerges---has paradoxically no counterpart in the analogous Reynolds number 
parameter for the most generic case of statistically homogeneous, isotropic inertial turbulence.

We perform detailed direct numerical simulations (Appendix A)  of the evolution equation
\begin{equation}
\partial_t {\bf u} + \lambda {\bf u}\cdot \nabla {\bf u}=-\nabla p - \Gamma_0\nabla^2 {\bf u} - \Gamma_2 \nabla^4 {\bf u} - (\alpha + \beta \vert {\bf u} \vert^2) {\bf u} \label{eq:genHyd}
\end{equation}
for the incompressible, coarse-grained velocity field ${\bf u}({\bf x},t)$ of the 
active, bacterial flow~\citep{wensink2012meso}. The nature---pusher or puller---of the constituent bacterium is determined by the sign of $\lambda$; 
our study focusses on pushers with $\lambda = 3.5$. In the Toner-Tu driving term~\citep{TT95,TT98}, $\beta > 0$ for stability and $\alpha < 0$ ensures an active injection of energy at scales $1/\sqrt{\vert\alpha\vert \beta}$. The coefficients  $\Gamma_0$ and $\Gamma_2$ lead to length $L_\Gamma=\sqrt{{\Gamma_2}/{\Gamma_0}}$ and time 
$\tau_\Gamma = {\Gamma_2}/{\Gamma_0^2}$ scales which arise from linear instabilities. We choose parameters consistent with experiments~\cite{wensink2012meso,joy2020friction,Sid21,james2021emergence}: $\Gamma_0 = 0.045$, $\Gamma_2 = \Gamma_0^3$, $\beta = 0.5$ and $-8 \leq \alpha \leq -1$ approximating flows with bacterial velocities in the range $25-75$ $\mu \rm{m/s}$.

\begin{figure*}
\includegraphics[width=0.75\linewidth]{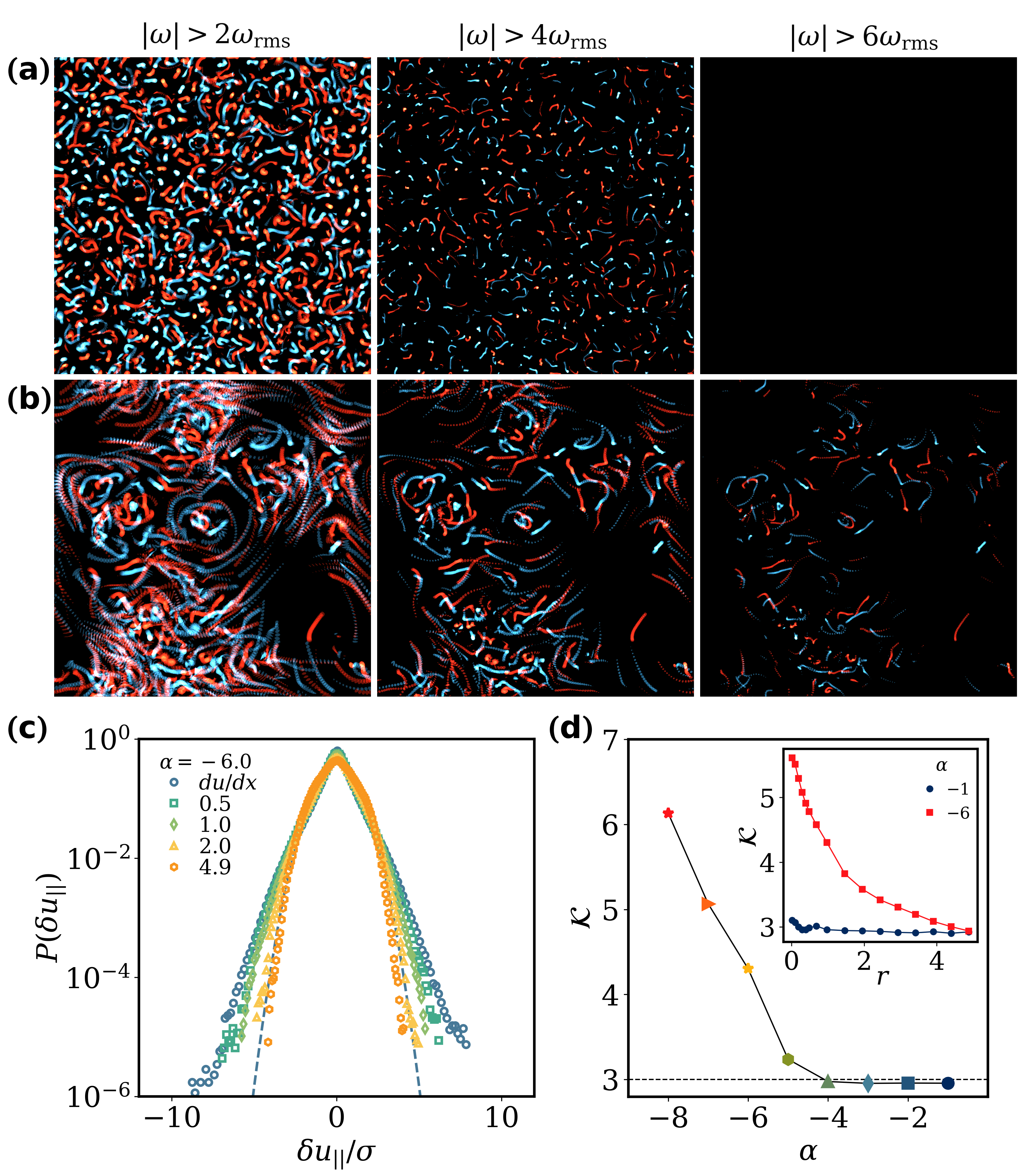}
	\caption{Vortex trails for (a) mildly ($\alpha=-1$) and (b) highly ($\alpha=-6$) active suspensions, shown as regions with vorticity above increasing thresholds, superimposed over 20 snapshots each separated by $\Delta t = 0.02$ (blue and red represent positive and negative values of $\omega$). While the mildly active case becomes quiescent at increasing thresholds, the highly active suspension shows spatio-temporal intermittency with large values of $\omega$ being prevalent. (c) Probability densities of the longitudinal velocity increments $\delta_r u$ (as well as the velocity gradient $du/dx$) 
		for different values of $r$ at $\alpha = -6$ with the Gaussian (dashed line) shown for comparison. 
	(d) The onset of intermittency is measured through the kurtosis as a function of scale $r$ 
	for different activities (inset); $\mathcal{K} > 3$ for a wide range of $r$ when $\alpha \lesssim \alpha_c$. This sharp transition with $\alpha$ shows up clearly in a plot of $\mathcal{K}$ as a function of $\alpha$ 
	(main panel) for a fixed $r$ (here $r = 1$).}
\label{fig:VelDiff}
\end{figure*}

Taking our cue from high Reynolds number turbulence, we begin investigating signatures
of intermittency, with increasing activity. Figs.~\ref{fig:VelDiff}(a) and~\ref{fig:VelDiff}(b) show vortex trails, as a temporal superposition of regions with the vorticity magnitude $|\omega|$ greater than multiples of their respective root-mean-square vorticity $\omega_{\rm rms}$. Low threshold vortex trails are equally prevalent for mildly ($\alpha=-1$) and highly ($\alpha=-6$) active suspensions. At higher thresholds ($|\omega| \geq 5\omega_{\rm rms}$), mildly active suspensions appear quiescent, whereas the highly active suspension continues to show strong deviations in $\omega$. We quantify this behaviour by considering the distribution of
(longitudinal) velocity increments $\delta_r u = [{\bf u}({\bf x} +
{\bf r}) - {\bf u}({\bf x})]\cdot\frac{{\bf r}}{|{\bf r}|}$.
In fully developed turbulence, the analogue of such measurements in the
so-called inertial range shows a strong departure from a Gaussian distribution
and the fat tails suggest that velocity increments are
intermittent with bursts of extreme values~\citep{Frisch-CUP}. For low activities, the
distributions of the increments, as well as the velocity gradients, are Gaussian, as also noted in
previous studies~\cite{wensink2012meso,bratanov2015new}. However, we find (Fig.~\ref{fig:VelDiff}(c))
that as $\alpha \lesssim -5$, the distributions become distinctly non-Gaussian, reminiscent of high Reynolds number turbulence~\cite{Frisch-CUP}.

This departure from a Gaussian is a sign of intermittency and is quantified by measuring the kurtosis $\mathcal{K} = \frac{\langle \delta_r u^4\rangle}{\langle  \delta_r u^2\rangle^2}$
as a function of $r$ as shown in the inset of Fig.~\ref{fig:VelDiff}(d). While for mild levels of activity ($\alpha = -1$) we see a scale-independent $\mathcal{K} = 3$ as it ought to be for Gaussian distributions, 
when $\alpha \lesssim -5$, clear evidence of intermittency appears as  
$\mathcal{K} > 3$ over a wide range of $r$. Fixing on a single (representative) value of $r = 1.0$, we see (Fig.~\ref{fig:VelDiff}(d)) a clear 
rise in $\mathcal{K}$ from the Gaussian limit as soon as $\alpha \lesssim -5$. 
This emergent intermittency suggests that active suspensions may well have asymptotic (in activity) states which have more 
in common with inertial turbulence than previously appreciated.

The transition from a non-intermittent to an intermittent flow has important
consequences for scale-invariance in such suspensions. A useful probe for this
is the energy spectrum $E(k)$ characterising the
(self-similar) distribution of kinetic energy across Fourier modes~\cite{Frisch-CUP}.  
The scaling form for the energy spectrum is easy to obtain dimensionally from 
the (advective) energy flux $\Pi(k)$ (Appendix B) and the local (scale-dependent) turnover
timescale $\tau_{\rm eff}(k)$ via $\tau_{\rm eff}(k)\Pi(k) \sim k E(k)$. At mild activity, velocity increments are Gaussian and the
effective timescale $\tau_{\rm eff}(k)$ is a constant, independent of $k$, as shown by Bratanov \textit{et al.}~\cite{bratanov2015new}. By using this
argument in the spectral equation, we obtain $E(k) \sim k^{\delta}$ with, unlike in inertial turbulence, 
an activity-dependent, \textit{non-universal} scaling exponent $\delta = \tau_{\rm eff}(2\alpha + 8\beta E_{\rm tot})/\lambda - 1 > 0$; where 
$E_{\rm tot} = u_{\rm rms}^2/2$~\cite{bratanov2015new,kiran2022irreversiblity}.

The assumption of a scale-independent $\tau_{\rm eff}$ is reasonable when
$\delta \gg 0$: Larger scales carry less energy and their role in distorting
small scales is marginal. However, $\delta$ decreases as a function of increasing activity and eventually $\underset{\alpha \to \alpha_c}{\lim} \delta \to 0$.
Consequently, this assumption must break down as larger scales become more
energetic at a cross-over or critical level of activity $\alpha_c$. A scale-dependent $\tau_{\rm eff}$, as we will show, also inevitably implies an $\alpha-$independent spectral scaling exponent $\delta$, which suggests a change in the nature of the flow as $\alpha \to \alpha_c$ to a possibly universal state. Naively,
setting $\delta = 0$,  we obtain $\alpha_c \equiv \lambda/2\tau_{\rm
eff} -2E_{\rm tot}$ (for $\beta = 0.5$). This of course makes the strong
assumption that $\tau_{\rm eff}$ is strictly scale-independent all the way up
to $\delta = 0$.

We know that $E_{\rm tot}$ itself is a function of $\alpha$ and, empirically as
long as $\alpha > \alpha_c$, the root-mean-squared velocity $u_{\rm rms} =
\sqrt{2E_{\rm tot}} \sim c_1\alpha + c_2$, where $c_1 < 0 $ and $c_2 > 0 $ are
constants~\cite{bratanov2015new,joy2020friction}. By using this expression, and solving the
resultant quadratic equation, 
we obtain $\alpha_c \approx -10$. The change in flow behaviour at $\alpha \approx - 5$ (Fig.~\ref{fig:VelDiff}(d) and in what follows), occurs well before this theoretical prediction. This hints that the assumption of a constant
$\tau_{\rm eff}$ becomes weak and indeed breaks down as $\alpha \to \alpha_c$ from above. 

\begin{figure}
\centering 
\includegraphics[width=0.98\linewidth]{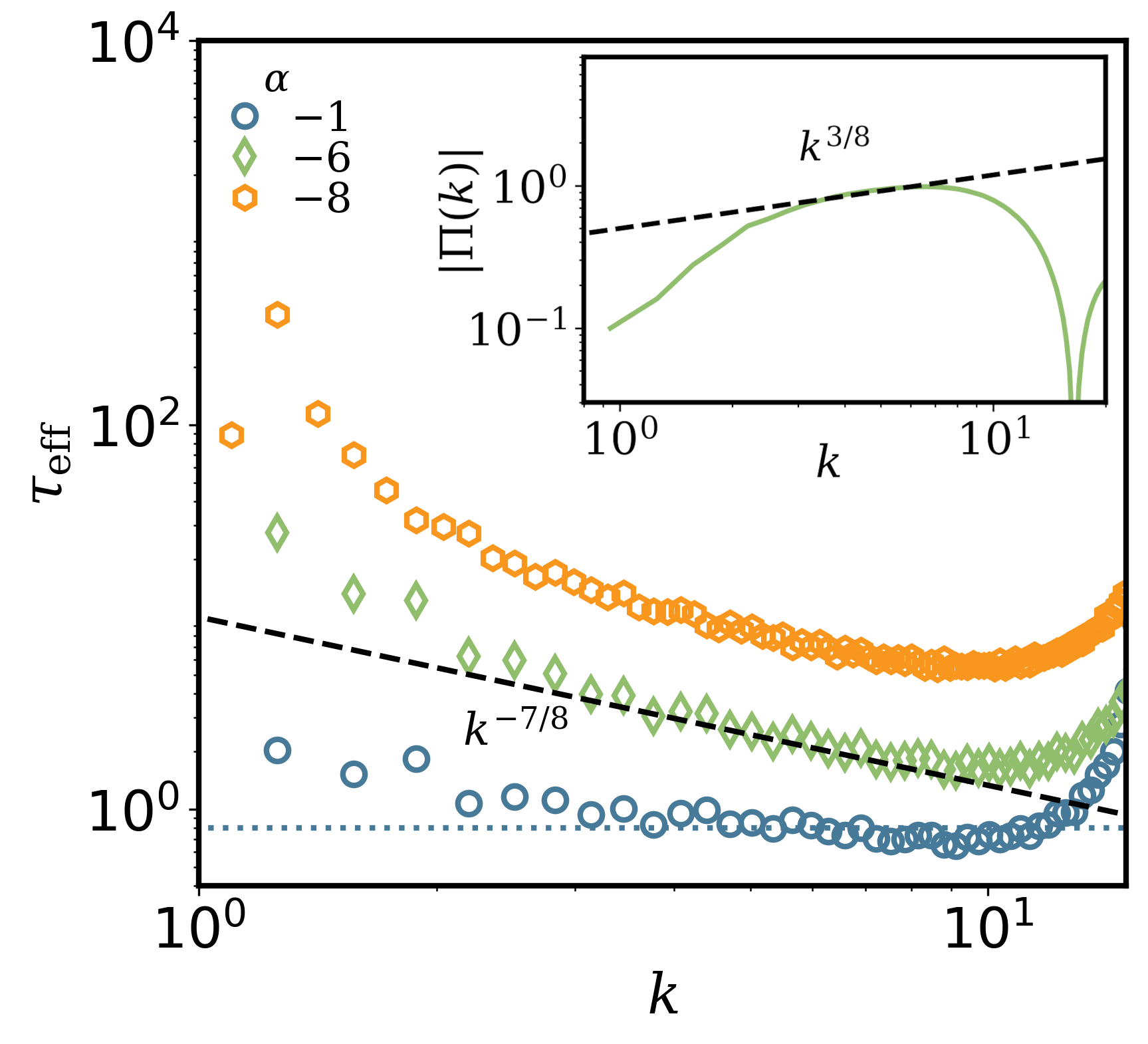}
	\caption{The transition from a scale-independent $\tau_{\rm eff} = {\rm const.}$ to a scale-dependent $\tau_{\rm eff} = k^{-7/8}$ as 
	the activity becomes stronger than $\alpha_c$ (curves vertically shifted for clarity). Inset: The advective flux for $\alpha=-6$, together with a $k^{3/8}$ scaling (dashed line).
	}
\label{fig:tau}
\end{figure}	


Assuming a scaling form $E(k) \sim k^{\delta}$, the local turnover time scale $\tau_{\rm eddy} \sim
(\sqrt{k^3E(k)})^{-1}$ can no longer be ignored as larger scales become more energetic when $\delta \lesssim 0$. Alongside, the flow reorganisation
seen in earlier studies of highly active turbulence ($\alpha \lesssim
\alpha_c$)~\cite{Sid21}, suggests an additional source of ``noise'' which accentuates
non-local interactions in Fourier space.  We conjecture a simple 
self-similar timescale $\tau_\alpha \sim 1/k$ for this noise, which may arise from a constant, activity induced velocity $v_0 = \sqrt{|\alpha|/\beta}$ acting across scales, leading
to an \textit{ansatz} $\tau_{\rm eff} = (\tau_{\rm eddy}\tau_\alpha)^{1/2} \sim k^{-\left (\frac{5+\delta}{4}\right )}$ of 
a scale-dependent effective time-scale for $\alpha \lesssim \alpha_c$.

In Fig.~\ref{fig:tau} we show a log-log plot of $\tau_{\rm eff} \equiv
\frac{kE(k)}{\Pi(k)}$ versus $k$ for $\alpha$ values on either side of  
$\alpha_c$. Consistent with the phenomenology of active turbulence at $\alpha
\gtrsim \alpha_c$, this effective time-scale is indeed a constant and
independent of $k$~\cite{bratanov2015new}. However, as soon as $\alpha \lesssim
\alpha_c$, a clear power-law $\tau_{\rm eff} \sim k^{-7/8}$ emerges.  This
implies, by using the preceding argument, that $E(k) \sim k^{-3/2}$ $\forall
\alpha \lesssim \alpha_c$.  

\begin{figure}
\centering 
\includegraphics[width=0.98\linewidth]{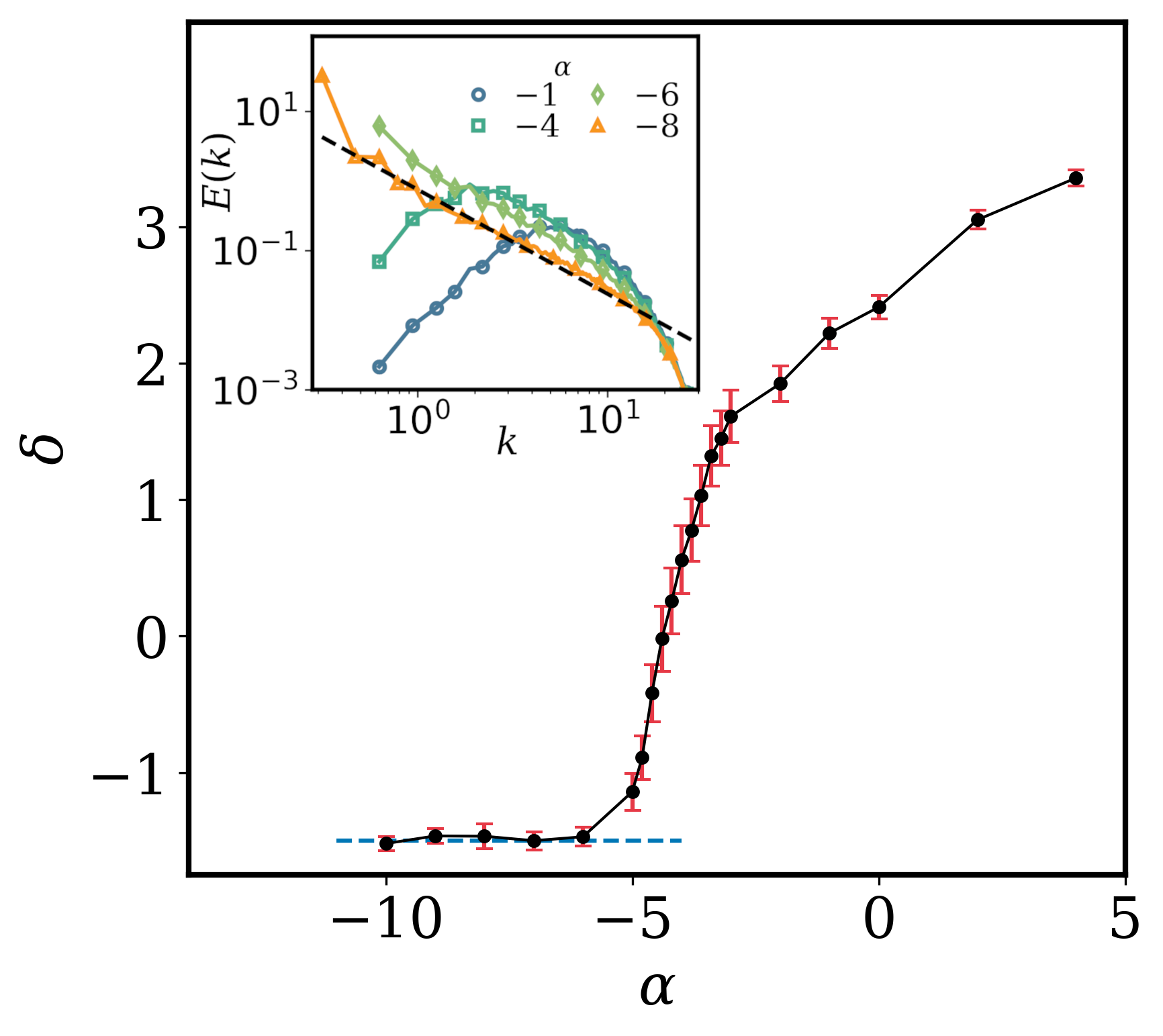}
	\caption{Inset: Log-log plots of the kinetic energy spectrum with increasing activity; the dashed black line shows a $k^{-3/2}$ scaling as a guide to the eye. The 
		spectral slopes $\delta$ (along with their error bars) as a function  of $\alpha$, in the main panel, show a sharp transition at a critical level of activity $\alpha_c \approx -5$ with the 
	$\delta = -3/2$ value denoted by the horizontal dashed line.
	}
\label{fig:Spectra}
\end{figure}	

We check this prediction for a universal scaling exponent $\delta = -3/2$ and, indeed, a new asymptotic phase of active turbulence for $\alpha \lesssim
\alpha_c$.  In Fig.~\ref{fig:Spectra} (inset) we show log-log
plots of the energy spectrum for different values of $\alpha$. Clearly, as long
as $\alpha \gtrsim \alpha_c$, the slope $\delta > 0$ decreases continually  and
$\delta \to 0$ as $\alpha \to \alpha_c$ from above.  For $\alpha \lesssim
\alpha_c$, the spectral slope seems to saturate, suggesting an emergent
universality, with $\delta = -3/2$.  This is clearly illustrated in Fig.~\ref{fig:Spectra} showing the scaling exponent $\delta$ (with
error bars) for different values of $\alpha$, extracted from plots such as
those shown in the inset. We find a sharp transition from a linear
(non-universal) dependence of the energy spectrum on activity, consistent with
earlier measurements~\cite{bratanov2015new,joy2020friction} to a constant,
universal asymptotic state $\delta = -3/2$ as $\alpha \lesssim \alpha_c \approx
-5$.

While we show a simple transition in active turbulence to an asymptotic,
universal intermittent state (Fig.~\ref{fig:VelDiff}) with a scale-dependent
time-scale (Fig.~\ref{fig:tau}) and a constant spectral exponent
(Fig.~\ref{fig:Spectra}), the precise value of $\delta$ cannot be derived independently. Therefore, to make our conjecture robust, we perform an additional self-consistency check by observing that the scaling forms of
$E(k)$ and $\tau_{\rm eff}$ for $\alpha \lesssim \alpha_c$ implies a scale-dependent flux $|\Pi(k)| \sim k^{3/8}$. 
(The lack of constant fluxes in such systems, unlike inertial turbulence, has been 
observed in previous studies~\cite{bratanov2015new,kiran2022irreversiblity} as well, although its precise 
form remained unobserved.) 
While scaling measurements of flux are less clean than for energy spectra, we find (Fig.~\ref{fig:tau}, inset) 
that the form of the flux is not inconsistent with the theoretical conjecture. 

\begin{figure*}
\centering
\includegraphics[width=\linewidth]{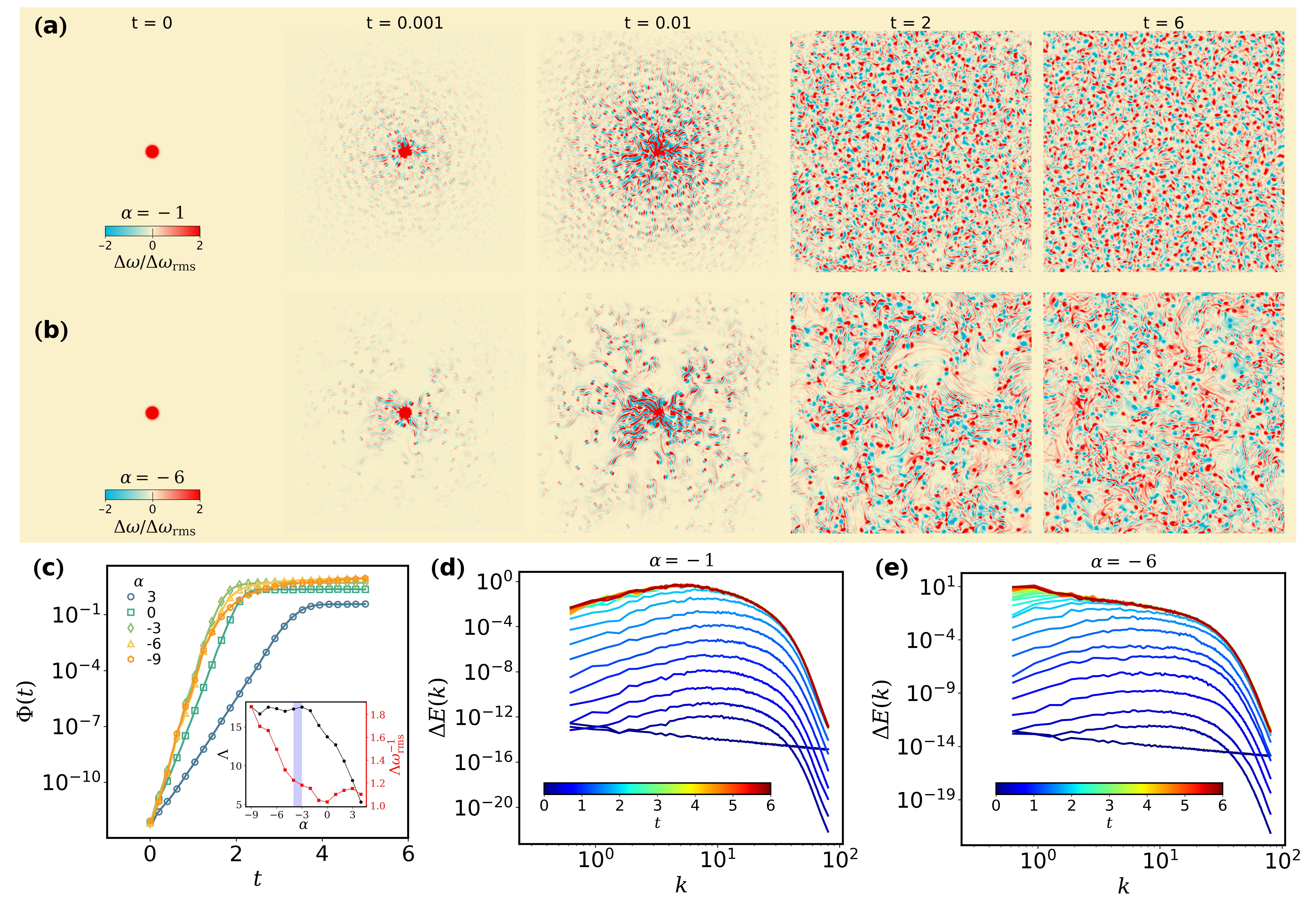}
	\caption{(a) Pseudo-color plots of an initially localised (at the center) perturbation $\Delta \omega ({\bf x},t)$, normalized by the instantaneous $\Delta \omega_{\rm rms} $, at different time instances for (a) 
	$\alpha=-1$ and (b) $\alpha = -6 \lesssim \alpha_c$ (see \url{https://www.youtube.com/watch?v=t0cLEbuuxhY} for the full evolution, made with Processing~\citep{reas2007processing,pearson2011generative}) (c) Semi-log plots of the decorrelator $\Phi(t)$ for different values of $\alpha$ showing an  
	exponential growth followed by saturation at a value $2E_{\rm tot}$. Inset: Plots of the Lyapunov (left axis) and the normalised Lyapunov (right axis) exponents as a function of $\alpha$. 
Log-log plots of $\Delta E(k)$ for (d) $\alpha = -1$ and (e) $\alpha = -6$ at different times. While both grow 
	exponentially, the $\alpha = -6$ case shows distinct multiscale features absent for 
	$\alpha = -1$.}	
\label{fig:Chaos}
\end{figure*}

The heuristic argument outlined above is physically appealing; nevertheless a
more rigorous, analytic way to show the transition at $\alpha \lesssim
\alpha_c$, thence $\tau_{\rm eff} \sim k^\xi$ with $\xi \neq 0$, to a universal
exponent $\delta = -3/2$ is by using the (approximate) equation of motion
(assuming closure at the level of the fourth moment) for
$E(k)$~\cite{bratanov2015new}. As detailed in Appendix B, one can then show that $\xi \neq 0$ leads inevitably to an
$\alpha$-independent $\delta$. Further, a solution of this spectral equation
yields an energy spectrum $E(k) \sim k^{-3/2}$ with an $\alpha$-dependent
exponential tail (Appendix B). The agreement between both the analytical and
phenomenological approaches, along with the compelling numerical evidence,
leaves little doubt about the existence and robustness of this critical
activity parameter.

While the lack of a constant flux is in sharp contrast to high Reynolds number
turbulent flows, the emergent state seems to share more in common
with inertial turbulence: Intermittency, non-Gaussianity and a universal
scaling of the energy spectrum. Inertial turbulence has another significant
attribute, namely, the dependence of chaos---quantified by a positive Lyapunov
exponent $\Lambda$---on the Reynolds number of the flow. This is particularly
interesting since the lack of a \textit{true} inertial range (characterized by
constant energy flux across scales) in active turbulence results in no
appreciable widening of the range of scales, with increasing activity, over which the
spectral scaling $k^{-3/2}$ holds. Thus, from this point of view, increasing the level of activity
is \textit{not analogous} to increasing Reynolds number in inertial turbulence. It is
reasonable to expect, therefore, that maximally saturated chaotic states and hence the 
efficacy of biomixing and transport in such bacterial suspensions, emerge around $\alpha \lesssim \alpha_c$ in sharp contrast to the uncurtailed
Reynolds dependent $\Lambda$ found in inertial
turbulence~\citep{mukherjee2016predictability,boffetta2017chaos,berera2018chaotic}.

In order to test these ideas, we set up perturbed twin-simulations ~\cite{mukherjee2016predictability,boffetta2017chaos,berera2018chaotic}, 
which allow tracking the evolution of a controlled initial (white noise) perturbation (Appendix C), and thence a measure of Eulerian chaos. The divergence between the unperturbed A and perturbed B solutions is quantified by the difference-vorticity $\Delta \omega ({\bf x},t)
\equiv \omega^{\rm B}({\bf x},t) - \omega^{\rm A} ({\bf x},t) $ and difference-velocity
$\Delta {\bf u} ({\bf x},t) \equiv {\bf u}^{\rm B}({\bf x},t) - {\bf u}^{\rm
A}({\bf x},t)$ fields, and the evolution of the perturbation is governed by the time-dependence of the decorrelator $\Phi(t) \equiv \langle \frac{1}{2}|\Delta {\bf u} ({\bf x},t)|^2\rangle$ (where $\langle\ \rangle$ denotes spatial and ensemble averaging)~\cite{Subhro2018,Sugan2020}

Understandably, at long times, systems ${\rm A}$
and ${\rm B}$ decorrelate and hence $\Phi(t)$ saturates to $2E_{\rm tot}$. At short times, however, we expect an exponential growth $\Phi(t) \sim
\exp{\Lambda t}$ indicative of the chaotic nature of these suspensions. Figure~\ref{fig:Chaos}(b) confirms these two behaviours for different $\alpha$, from which we extract the Lyapunov exponent $\Lambda$---a measure
of the level of chaos in the system---and examine its dependence on activity (Fig.~\ref{fig:Chaos}(b), inset; left axis).
Interestingly, and perhaps unsurprisingly, $\Lambda$ increases 
monotonically with activity and achieves a maximum as $\alpha \to
\alpha_c$ (from above) and then plateaus. Thus, active suspensions are indeed maximally chaotic, and remain so, as $\alpha \lesssim \alpha_c$ (indicated by the vertical blue bar). 
The significance of these maximally chaotic states is best understood by normalizing $\Lambda$ with $\omega_{\rm rms}$. A plot of this 
normalised Lyapunov exponent (Fig.~\ref{fig:Chaos}(b), inset; right axis) reveals that the perturbation growth timescale, in fact, becomes smaller in comparison to the vortex timescale, when $\alpha \lesssim \alpha_c$. 
In other words, the chaoticity of the suspension increases fundamentally when $\alpha$ goes beyond $\alpha_c$, 
and not as a consequence of more vigorous advection. 

A visual impression of how these systems decorrelate is best obtained by introducing a localized perturbation. 
In Figs~\ref{fig:Chaos}(a) and~\ref{fig:Chaos}(b) we show representative pseudo-color plots of $\Delta \omega$ for $\alpha  = -1$ and $\alpha  = -6 < \alpha_c$, respectively, over time. The initially ($t = 0$) localised Gaussian ($\sigma=0.02L$) perturbation spreads rather quickly in a self-similar manner till the $\Delta \omega$ field itself becomes 
indistinguishable from the corresponding vorticity fields (Appendix C). (The Lyapunov exponents measured from 
such spatially localised perturbations are consistent with those shown in Fig.~\ref{fig:Chaos}(a), and these results are qualitatively insensitive to the exact nature and amplitude of the perturbation.)

Finally, we examine the spectral growth of the perturbation energy by tracking the energy spectrum $\Delta E(k)$ of the difference-velocity field $\Delta {\bf u}$. Figure~\ref{fig:Chaos}(c) shows that, for $\alpha=-1$, the initial perturbation rapidly assumes a self-similar spectral shape which grows at a constant exponential rate (equidistant curves along the vertical log-scale) until saturation, revealing a single dominant Lyapunov exponent since the mildly active flow has a single vortex scale. At $\alpha=-6$, the flow becomes truly multiscale, which is reflected in Fig.~\ref{fig:Chaos}(d) where the initial perturbation first assumes a self-similar shape till the saturation of the high wavenumbers, followed by a successive (and slower) saturation of the low wavenumbers. We highlight that this behaviour is reminiscent of inertial turbulence~\citep{mukherjee2016predictability}, where the growth of perturbation energy slows down as successively larger scales saturate, \`a la Lorenz~\citep{lorenz1969predictability}. However, the saturation of $\Lambda$ with increasing activity is unlike the unbounded growth of $\Lambda$ with Reynolds number in inertial turbulence~\citep{mukherjee2016predictability,boffetta2017chaos,berera2018chaotic}.

In this paper we provide evidence that low Reynolds number active flows, beyond a
critical threshold of activity $\alpha_c$, are universal in a manner similar to inertial turbulence. 
Interestingly, unlike the case
of inertial, homogeneous and isotropic turbulence where the estimate of a
critical Reynolds number is moot, active suspensions seems to allow a  critical
value of activity when a truly turbulent and universal phase emerges. 
This is summarised in terms of a transition, 
most dramatically seen in the spectral exponent: 
\begin{equation}
  \delta =
    \begin{cases}
	    \frac {\tau_{\rm eff}(2\alpha + 8\beta E_{\rm tot})}{\lambda} - 1\gtrsim 0 & \text{$\alpha \gtrsim \alpha_c$ ~\cite{bratanov2015new}}
    \\
    -3/2 & \text{$\alpha \lesssim \alpha_c$}
  \end{cases}
  \label{eq:delta}
\end{equation}

While the evidence for this transition at $\alpha_c \approx -5$ leaves little room 
for doubt, the form of the scaling is reminiscent of 
several classes of turbulent flows where $\delta = -3/2$. The most well-known example 
of this is perhaps magnetohydrodynamic turbulence~\cite{Iroshnikov,Kraichnan} but other instances are known in 
the area of wave turbulence~\cite{Nazarenko-review}--such as acoustic turblence~\cite{Zakharov}--active binary fluid turbulence~\citep{pan2022energy}, as well as more unrelated examples
like the Burgers equation on a fractally disordered (Fourier) lattice~\cite{decimated-Burgers}. It is important in future work to 
understand if active turbulence in the $\alpha \lesssim \alpha_c$ regime can be described in terms of the formalisms developed 
in these areas. We believe that experiments on extremely active suspensions, for instance involving bacteria that swim an order of magnitude faster~\cite{petroff2015fast}, would also shed light 
on the precise nature of the turbulence displayed in such asymptotic states. We should underline, however, that the 
existence of a cross-over activity $\alpha_c$ is independent of the 
precise form in which the energy spectrum scales. Furthermore, the emergence of a maximally chaotic state in active turbulence again shows that a broad-brushed parallel with inertial turbulence obfuscates rich phenomenology, all of which must be tied intimately with biologically relevant strategies for survival and growth.

\begin{acknowledgements} 
The simulations were performed on the ICTS clusters
{\it Tetris} and {\it Contra}. SM and SSR thank J\'er\'emie Bec, Sugan D. Murugan, and Jason Picardo  for insightful discussions and suggestions. SSR acknowledges
	SERB-DST (India) projects MTR/2019/001553, STR/2021/000023 and CRG/2021/002766 for
financial support. MJ gratefully acknowledges support from the grant STR/2021/000023 and the hospitality of 
	ICTS-TIFR. SSR would like to thank the Isaac Newton Institute for Mathematical Sciences for support and hospitality 
	during the program \textit{Mathematical aspects of turbulence: where do we stand?} (EPSRC Grant Number EP/R014604/1) when a part of this work was done.  The authors acknowledge the support of the DAE, Govt. of
India, under project no.  12-R\&D-TFR-5.10-1100 and project no.  RTI4001.
\end{acknowledgements}

\appendix 

\section{Appendix A: Direct Numerical Simulations}
We numerically integrate the generalized Navier-Stokes equation~\eqref{eq:genHyd}
on a periodic (square) domain, by using a fully dealiased pseudo-spectral method~\citep{james2018vortex,james2018turbulence}. We use various domain sizes in the range $20 \leq L \leq 80$ with up to $1024 \leq N_x \leq 4096$ collocation points. Simulations are performed with a time-step of $dt = 0.0002$ for upto $5\times 10^5$ iterations, beyond an initial spin-up time of $20000$ iterations to reach a statistically steady-state. Spectra and fluxes are written frequently to perform ensemble averaging over around $10000$ samples, while velocity increment statistics are averaged over upto $50$ field snapshots. Note that simulations of highly active turbulence, $\alpha \leq -7$ are performed on $L=80$ ($4096^2$) domains, to be able to resolve the larger structures without forming condensates.\\

\section{Appendix B: Spectral Analysis: Critical Activity and Universal Scaling Expoent}
Each term in Eq.~\eqref{eq:genHyd} can be decomposed into its Fourier series, where for instance the velocity, ignoring time dependence, becomes
\begin{equation}
{\bf u}({\bf r}) = \sum_{\bf k} {\bf u}({\bf k})e^{i{\bf k}\cdot {\bf r}}
\end{equation}
where ${\bf k}$ has components $k_x$ and $k_y$ that are integer multiples, $\left\lbrace 1,2...n\right\rbrace$, of $2\pi/L$ where $L$ is the physical system size along one-direction and $n=N_x/2$ where $N_x$ is the number of collocation points along one-direction. The Fourier coefficients are given as
\begin{equation}
{\bf u}({\bf k}) = \frac{1}{L^2} \int_0^L\int_0^L {\bf u}({\bf r})e^{-i{\bf k}\cdot {\bf r}}{\rm d}x{\rm d}y
\end{equation}
Further, the energy spectrum over the scalar wavenumber $k$ is defined as
\begin{equation}
E(k) = \frac{1}{2}\sum_{k^\prime=k-1/2}^{k + 1/2} \left \langle {\bf u}({\bf k^\prime})\cdot {\bf u}({\bf k^\prime}) \right \rangle
\label{eq:EnergySpectrumDefn}
\end{equation}
where $k = \sqrt{{\bf k}\cdot {\bf k}}$ and $\left\langle \right\rangle$ denotes ensemble averaging. Similarly, taking the Fourier transform of the terms of Eq.~\eqref{eq:genHyd}, multiplying the resultant equation with ${\bf u}^\star({\bf k})$ (which is the complex-conjugate of ${\bf u}({\bf k})$) and taking the shell-sum as in Eq.~\eqref{eq:EnergySpectrumDefn} gives the energy spectrum equation~\cite{bratanov2015new,kiran2022irreversiblity} as
\begin{equation}
\frac{\partial E(k)}{\partial t} = 2\gamma(k) E(k) - \lambda T^{\rm adv}(k) + T^{\rm cub}(k)
\label{eq:ek}
\end{equation}
where $T^{\rm adv}(k)$ and $T^{\rm cub}(k)$ are the advective and cubic terms, respectively, and $T^{\rm adv}(k)$ appears with a pre-factor $\lambda$ due to the generalized non-linear term in Eq.~\eqref{eq:genHyd}. Here $\gamma(k)$ is the spectral form of the linear terms in Eq.~\eqref{eq:genHyd}, and is given as $\gamma(k) = -\alpha + \Gamma_0 k^2 - \Gamma_2 k^4$. Further, the energy flux is defined as $\Pi(k) = -\int_0^k T^{\rm adv}(p) {\rm d}p$, which can be dimensionally related to the energy spectrum, using a scale-dependent effective timescale $\tau_{\rm eff}$ as 
\begin{equation}
\tau_{\rm eff} \Pi(k) \equiv \lambda kE(k)
\label{eq:FluxEkRelation}
\end{equation}
Using a general form of $\tau_{\rm eff} = k^\xi/c$ (where $c$ is some dimensional constant), the flux can be re-written as
\begin{equation}
\Pi(k) \equiv c\lambda k^{1-\xi} E(k)
\label{eq:FluxTerm}
\end{equation}
Following \citet{bratanov2015new}, we use the quasi-normal approximation $T^{\rm cub}(k) \approx -8\beta E_{\rm tot}E(k)$ where $E_{\rm tot} = u_{\rm rms}^2$, which for a statistically stationary state reduces Eq.~\eqref{eq:ek} to 
\begin{equation}
-2(\alpha + 4\beta E_{\rm tot} - \Gamma_0k^2 + \Gamma_2k^4)E(k)+ \frac{{\rm d}\Pi(k)}{{\rm d}k} = 0
\label{eq:ode}
\end{equation}
At mild levels of activity, where $\tau_{\rm eff} = {\rm const.}$ ($\xi = 0$)~\citep{bratanov2015new} in Eq.~\eqref{eq:FluxEkRelation}, one simply gets $\Pi(k) = \lambda kE(k)/\tau_{\rm eff}$, and ignoring the $\Gamma_2$ term (at low wavenumbers), Eq.~\eqref{eq:ode} can be solved to obtain the energy spectrum scaling~\citep{bratanov2015new} as
\begin{equation}
E(k) = \widetilde{E}_0 k^\delta \exp\left( -\frac{\Gamma_0\tau_{\rm eff}}{\lambda}k^2 \right)
\end{equation}
where $\delta = (2\alpha + 8\beta E_{\rm tot})\tau_{\rm eff}/\lambda -1$, with $\widetilde{E}_0$ a constant of integration. The spectral slope in this mildly active regime, where $\tau_{\rm eff} = {\rm const.}$ ($\xi = 0$), varies with $\alpha$, as also observed in the simulations.

However, for highly active suspensions, with a scale-dependent $\tau_{\rm eff} = k^\xi/c$ ($\xi \neq 0$ for the reasons explained in the main text) we retain the general form of $\Pi(k) = c\lambda k^{1-\xi} E(k)$. By using this, Eq.~\eqref{eq:ode} leads to (again, ignoring the $\Gamma_2$ term)
\begin{align}
-2(\alpha + 4\beta E_{\rm tot} - \Gamma_0k^2)E(k) &+ c\lambda(1-\xi)k^{-\xi} E(k) \nonumber \\
		&+ c\lambda k^{1-\xi} \frac{{\rm d}E(k)}{{\rm d}k} = 0
\end{align}
which we rearrange as
\begin{equation}
\frac{{\rm d}E(k)}{E(k)} = \frac{2(\alpha + 4\beta E_{\rm tot})}{c\lambda} k^{\xi-1}{\rm d}k - \frac{2\Gamma_0}{c\lambda} k^{\xi+1} {\rm d}k -(1-\xi) k^{-1}{\rm d}k
\label{eq:GeneralForm}
\end{equation}
%
Equation~\eqref{eq:GeneralForm} can be solved to obtain the energy spectrum as
\begin{equation}
E(k) = \widetilde{E}_0 k^{\xi-1} \exp\left( \frac{2(\alpha + 4\beta E_{\rm tot}) k^{\xi}}{c\lambda\xi} - \frac{2\Gamma_0 k^{\xi+2}}{c\lambda(\xi+2)} \right) 
\label{eq:AnalyticalSpectrum}
\end{equation}
Importantly, when $\xi > 0$ (i.e. when $\tau_{\rm eff}$ becomes scale dependent), the theoretical energy spectral scaling also becomes $\alpha$ independent at low $k$ (since the first term on the RHS of Eq.~\eqref{eq:GeneralForm} 
integrates to algebraic instead of logarithmic as in the case of $\xi=0$), hence consistent with the empirical observation of an $\alpha$-independent asymptotic universal scaling in the numerical simulations. 
By using the observed scaling of $\tau_{\rm eff} \sim k^{-7/8}$, we get
\begin{equation}
E(k) = \widetilde{E}_0k^{-15/8} \exp\left( \frac{2(\alpha + 4\beta E_{\rm tot}) k^{-7/8}}{-7c\lambda/8} - \frac{2\Gamma_0 k^{9/8}}{9c\lambda/8} \right) 
\label{eq:AnalyticalSpectrumScale}
\end{equation}
Finally,  $\widetilde{E}_0$ is still undetermined. In the analogous calculation for classical, inertial turbulence, because $\tau_{\rm eff} = \epsilon^{-1/3}k^{-2/3}$, where $\epsilon$ is the (constant) rate of energy dissipation, 
it can be shown that $\widetilde{E}_0 = \epsilon^{-2/3}$. However, in active turbulence, with a scale-dependent flux and no real scale-separation, such a straightforward calculation is difficult. Nevertheless, we can show by  
comparing the non-exponential part of Eq.~\eqref{eq:AnalyticalSpectrumScale} to the flux and energy spectrum relation of Eq.~\eqref{eq:FluxEkRelation} (with $\xi = -7/8$) that $\pik = c\lambda k^{15/8}\widetilde{E}_0 k^{-15/8}$, 
or $\widetilde{E}_0 = \pik/c\lambda$. Since $\pik$ is shown to have a weak scale invariance $\pik \sim k^{3/8}$, the spectral analysis yields (ignoring the exponential tail) 
\begin{equation}
E(k) \sim k^{-3/2}
\end{equation}
consistent with the numerically observed and phenomenologically explained (main text of the manuscript) $\alpha-$independent scaling of the energy spectrum as $\alpha \lesssim \alpha_c$. 
%
%

\section{Appendix C: Twin Simulations}
We choose an arbitrary
realisation of the statistically steady vorticity field $\omega^{\rm A}_0 =
\omega$, from the numerical solutions of Eq.~\eqref{eq:genHyd} for a given set
of parameters, and obtain $\omega^{\rm B}_0 = \omega^{\rm A}_0 + \delta
\omega_0\eta$. The two vorticity fields, denoted by superscripts ${\rm A}$ and $\rm
B$, are thus nearly identical up to a small perturbation $\delta \omega_0\eta$ introduced at each 
grid point at $t=0$. We fix the amplitude $\delta \omega_0 = 10^{-3}$ and $\eta \in [-1,1]$ is a uniformly distributed random noise. We use $\omega^{\rm A}_0$ and $\omega^{\rm B}_0$ as initial
conditions for simultaneous simulations of systems ${\rm A}$ and $\rm B$ and
measure, point-wise, the difference-vorticity $\Delta \omega ({\bf x},t)
\equiv \omega^{\rm B}({\bf x},t) - \omega^{\rm A} ({\bf x},t) $ and difference-velocity
$\Delta {\bf u} ({\bf x},t) \equiv {\bf u}^{\rm B}({\bf x},t) - {\bf u}^{\rm
A}({\bf x},t)$ fields as a function of space and time. The decorrelator $\Phi(t) \equiv \langle \frac{1}{2}|\Delta {\bf u} ({\bf x},t)|^2\rangle$ is obtained by ensemble averaging over space, as well as over 10 independent twin simulations. For visualization alone, we perform separate twin-simulations with spatially localized Gaussian perturbations as shown in the manuscript and in the supplementary movie (see also \url{https://www.youtube.com/watch?v=t0cLEbuuxhY}).\\

\bibliography{references}

\end{document}